\documentclass[prd,aps,showpacs,preprintnumbers,amssymb]{revtex4}
\usepackage{axodraw}
\usepackage{color}
\usepackage{epsf}

\def\e3p{$\eta \rightarrow 3 \pi$}

\begin{document}
\title{%
\hfill{\normalsize\vbox{%
\hbox{}
 }}\\
{About the particle structure and content of the standard model}}

\author{Renata Jora
$^{\it \bf a}$~\footnote[2]{Email:
 rjora@theory.nipne.ro}}

\affiliation{$^{\bf \it b}$ National Institute of Physics and Nuclear Engineering PO Box MG-6, Bucharest-Magurele, Romania}

\date{\today}

\begin{abstract}
We determine the number and distribution of the fermion states in the standard model based on the  possible fermion representations of the gauge bosons. By extracting the even parity scalars from  the fermion states we suggest the existence of multiple Higgses arranged in suitable singlets or multiplets of $SU(3)_L$ and  $SU(3)_c$.
\end{abstract}
\maketitle

The standard model of elementary particles \cite{Glashow}-\cite{Veltman} contains $12$ gauge bosons corresponding to the group $U(1)_Y\times SU(2)_L \times SU(3)_c$, one complex Higgs doublet, and three generations of fermions each containing $15$ right
handed or left handed fermion states. While one generation makes perfect sense from the point of view of symmetries involved and also for the cancellation of the anomalies there is no definite answer of why the number of generations should be multiplied by three. One can construct more sophisticated theories with enlarged symmetries to accommodate the proliferation of fermion states but up to now there have not been any experimental hint or proof that this indeed correspond to the reality. In \cite{Jora} we made our first attempt to clarify this problem.  Here we will propose a more comprehensive solution that relies only partially on what we claimed there.

We shall consider as it is often accustomed in the literature the gauge symmetry as the starting point of our discussion. The associated gauge bosons are: $B_{\mu}$ of $U(1)_Y$ with no definite parity, $A^a_{\mu}$ of $SU(2)_L$ which is left handed (if one regards this in terms of fermion states) and $G^b_{\mu}$ of $SU(3)_C$ which has the intrinsic parity $P=-1$ and it is a vector.

Since $B_{\mu}$ has no definite parity and it is a combination of both vector and axial vector it has the most general representation in terms of fermion states as:
\begin{eqnarray}
B_{\mu}=c_1\bar{\Psi}_L\gamma^{\mu}\chi_L+c_2\bar{\eta}_R\gamma^{\mu}\xi_R+h.c.
\label{res5534435}
\end{eqnarray}
Thus there are $4$ left or right handed fermion states that correspond to $B_{\mu}$.  This is in a sense both a minimal and a maximal representation; it is minimal because one might write this as linear combination of a vector and an axial vector with the same counting of the states and it is maximal because we do not repeat  in the combination any field.  We thus suggest that for any boson of the standard model there are fermion composites with the correct quantum numbers but not bound that might pop up from the vacuum such that both representations of the gauge boson survive. An equivalent statement is: Any change of variable with the property outlined in the previous sentences that can be made corresponds to degrees of freedom that have counterpart in the physical reality.

Next $A^a_{\mu}$ is definitely represented in terms of left handed fermion states. The correct representation is given by:
\begin{eqnarray}
(t^aA^a_{\mu})_{ij}=\bar{\xi}_{ij L}\gamma^{\mu}\zeta_{ijL}+h.c,
\label{res5546}
\end{eqnarray}
where $t^a$ is the generator of the group $SU(2)_L$ in the fundamental representation and the indices $i$ and $j$ can take the values $1$, $2$. The double index counts four left handed fermion states for each $\xi$ or $\zeta$  so in total there are  $8$ left  handed corresponding states. Knowing that that right handed field may be represented as the hermitian conjugate of an left handed field  the separation of states in terms of left handed and right handed fields should be considered only apparent. In the end what it is important is the number of degrees of freedom. In total we count for the number of chiral fermion states in the electroweak sector: $4+8=12$ which represents the number of chiral leptons if one assumes that there are also right handed neutrino states.

Applying a similar approach one can represent the vector color field $G^b_{\mu}$ as:
\begin{eqnarray}
(t^a_cG^a_{\mu})_{ij}=\bar{\eta}_{ij}\gamma^{\mu}\tau_{ij}+h.c.
\label{fin56657}
\end{eqnarray}
Here $i,j=1..3$ and in total they count $3\times 3\times 2=18$ ($3\times3=18$ from the double index $ij$ and $2$ because there are two distinct fermions $\eta$ and $\tau$)  Dirac fermion states. In terms of left handed and right handed
 states there are $36$ spinors.  Note that $18$ corresponds exactly to the number of Dirac quarks in the standard model.

If we add up all the chiral states representing in a consistent way the standard modes gauge bosons we obtain:
\begin{eqnarray}
4 (B_{\mu})+8 (A^a_{\mu})+36 (G^a_{\mu})=48
\label{tot7758}
\end{eqnarray}
corresponding to the exact number of left handed or right handed states in the standard model if one includes right handed neutrinos. Due to numerous experiments that probe the existence of neutrino masse and mixings the presence of neutrino right handed states is very probable. Moreover the results in Eq. (\ref{tot7758}) count exactly the number of Dirac leptons $6$ and quarks $18$ in the standard model. The actual fermion gauge eigenstates in the standard model might appear as any linear combinations of them and it is these linear combinations that are arranged  in the corresponding multiplets that we know. Thus the method cannot reveal the actual interaction terms of the fermions with the gauge bosons.

Our approach predicts the number and distribution of fermion states. But if it is indeed meaningful it can make also predictions in another direction: that of the number of scalar states.
We cannot extract scalar states from the gauge bosons because these would exhaust completely the degree of freedom of the gauge bosons. But we can apply the previous procedure this time in terms of the fermion states.  We start by considering the possible number of composite scalar states made of two chiral fermions $\Psi_L$  and $\Psi_R$ of the same species:
\begin{eqnarray}
&&\Psi_L^{\dagger}\Psi_R
\nonumber\\
&&\Psi_L^T\sigma^2\Psi_L
\nonumber\\
&&\Psi_R^T\sigma^2\Psi_R
\label{res554677}
\end{eqnarray}
These states would correspond to three complex scalars. We are interested in states with parity $P=1$ for reasons that will be outlined below. If we consider how the parity operator acts on a Dirac fermion $P\Psi(x)P=\gamma^0\Psi(t,-\vec{x})$ we deduce that from the states in Eq. (\ref{res554677}) only the following ones have the desired property:
\begin{eqnarray}
&&\Psi_L^{\dagger}\Psi_R+h.c.=\Phi_1(x)
\nonumber\\
&&\Psi_L^T\sigma^2\Psi_L+\Psi_R^T\sigma^2\Psi_R=\Phi(x).
\label{st5546}
\end{eqnarray}
These correspond to three real scalar degrees of freedom.
Consider a left handed spinor with the structure:
\begin{eqnarray}
\Psi_L(x)=\left(
\begin{array}{c}
\Psi_1(x)\\
\Psi_2(x)
\end{array}
\right).
\label{st554678}
\end{eqnarray}
We shall reexpress the spinor $\Psi_L(x)$ in terms of a new set of variables $\Phi(x)$, $\Psi_1'(x)$ and $\Psi_2'(x)$ as in:
\begin{eqnarray}
\Psi_L(x)=\Phi(x)\left(
\begin{array}{c}
\Psi_1'(x)\\
\Psi_2'(x)
\end{array}
\right)=\Phi(x)\Psi_L',
\label{res44232}
\end{eqnarray}
where $\Phi(x)=\Psi_L^T\sigma^2\Psi_L+\Psi_R^T\sigma^2\Psi_R$, $\Psi_1'(x)=\frac{\Psi_1(x)}{\Phi(x)}$ and $\Psi_2'(x)=\frac{\Psi_2(x)}{\Phi(x)}$. We can make a similar substitution for both the right handed spinor (with the same $\Phi(x)$ and for the full Dirac field (with $\Phi_1(x)$). Then the parity operator acts as follows:
\begin{eqnarray}
P\Psi_LP=\Psi_R(t,-\vec{x})=P\Phi(x)PP\Psi_L'P=\Phi(t,-\vec{x})P\Psi_L'P=\Phi(t,-\vec{x})\Psi_R'(t,-\vec{x}),
\label{rest456}
\end{eqnarray}
so it acts on the new state $\Psi_L'$ as on a regular spinor.
Note that if we had considered $\Phi(x)$ with mixed parity  this would have spoiled the parity properties of the spinor $\Psi_L'$. We thus conclude that we can extract three real scalars (or that there are two possibilities outlined in Eq. (\ref{st5546})) upon reparametrization from one Dirac fermion without spoiling the correct properties of the leftover spinor state.

Considering the $24$ full Dirac fermions existent in the standard model (if we admit the existence of right handed neutrinos) we count in total $72$ real Lorentz scalars. We suggest that these scalars with various structures and combined in various multiplets   should really be present in the theory. In \cite{Jora1} we introduce a criterion of consistency that claims that any theory at zero temperature must have the un-normalized partition function independent of momenta.  Thus  a match must exist between the fermion and boson degrees of freedom. The approach considered there suggests that there is  a gap in the partition function of $(p^2)^{36}$ ( if we exclude the Higgs boson) which can be annihilated only by the presence of additional $72$ real bosonic degrees of freedom including the standard model Higgs doublet. This confirms  the correctitude of our findings here.

The next step is to make arrangements in the singlets or multiplets of $SU(2)_L$ and $SU(3)_c$ for $72$ scalars. There is a large number of possibilities.  First we need to estimate how many scalars have color degree of freedom. The total number of real scalars associated with the quarks is $18\times 3=54$.  We know that each quark has three color degrees of freedom and this is true also for the corresponding scalars $\Phi_i(x)$ and $\Phi_{1i}$. However the sums $\Phi_i\Phi_i$ and $\Phi_{1i}\Phi_{1i}$ are colorless. Then for each three color degrees of freedom there is a colorless scalar and thus the number of colorless scalar degrees of freedom associated with the quark sector is $18$. We are left with $36$  real scalars that have color index.

We analyze first   the colored scalars. We can arrange these in color triplets, sextet or octets with respect to the group $SU(3)_C$. Furthermore each of them can be either singlets or triplets with respect to $SU(2)_L$. The color triplet \cite{Hagiwara}, \cite{Vecchi}  can be singlet of $SU(2)_L$ and thus has $6$ real degrees of freedom or triplet under $SU(2)_L$ with $18$ real degrees of freedom. The scalar sextet \cite{Hagiwara}can also be singlet under $SU(2)_L$  and thus have $12$ degrees of freedom or triplet under $SU(2)_L$ with $36$ degrees of freedom. The color octet \cite{Dobrescu1}, \cite{Dobrescu2} which is  singlet under $SU(2)_L$ has $8$ degrees of freedom whereas the octet triplet has $24$. We can drop the latter two along with other possible representation as they cannot account in any combination for the $36$ real degrees of freedom desired. Considering also the possible couplings with the quarks and eliminating the unnecessary repetition of the same type of scalars we are left with only one natural choice; the color sextet triplet. The colorless $18$ degrees of freedom can be saturated by  for example  three Higgs doublets ($12$ degree of freedom), each one corresponding to a generation and a $Y=-\frac{1}{3}$ Higgs triplet \cite{Vecchi}  of $SU(2)_L$ ($6$ real degrees of freedom).  By analogy the $18$ scalar degrees of freedom in the lepton sector will correspond to another three Higgs doublets, each acting on a generation of leptons and an additional $Y=2$ Higgs triplet \cite{Georgi}, \cite{Gunion}.

The final Lagrangian we propose has thus the form:
\begin{eqnarray}
{\cal L}={\cal L}_1+{\cal L}_2+{\cal L}_3+{\cal L}_4+{\cal L}_5.
\label{res43553}
\end{eqnarray}
Here ${\cal L}_1$ is:
\begin{eqnarray}
{\cal L}_1=-\frac{1}{4}B^{\mu\nu}B_{\mu\nu}-\frac{1}{4}F^{\mu\nu i}F^i_{\mu\nu}-\frac{1}{4}G^{\mu\nu a}F^a_{\mu\nu},
\label{res4355}
\end{eqnarray}
where the first kinetic term corresponds to the $U(1)_Y$ group, the second to $SU(2)_L$ and the third to $SU(3)_c$.
Then,
\begin{eqnarray}
{\cal L}_2=\bar{\Psi}i\gamma^{\mu}D_{\mu}\Psi,
\label{rew3232}
\end{eqnarray}
where $\Psi$ represent all standard model fermion fields and $D_{\mu}$ the covariant derivatives associated to them, Next we have the Higgses kinetic terms incorporated in ${\cal L}_3$:
\begin{eqnarray}
{\cal L}_3=\sum_{i=1}^3\sum_{j=1}^2(D^{\mu}H_i^j)^{\dagger} (D_{\mu} H_i^j)+\frac{1}{2}\sum_{i=1}^2 {\rm Tr}(D_{\mu}\Phi_i)^{\dagger}(D_{\mu}\Phi_i)+
{\rm Tr}(D^{\mu}\Sigma)^{\dagger}(D_{\mu}\Sigma).
\label{res43435}
\end{eqnarray}
The six complex Higgs doublets have the form:
\begin{eqnarray}
H_i^j=\left(
\begin{array}{c}
H_i^{j\dagger}\\
H^{0j}_i
\end{array}
\right).
\label{res5344398}
\end{eqnarray}
and the standard covariant derivatives. The $Y=-\frac{1}{3} $ Higgs triplet \cite{Vecchi} has the representation:
\begin{eqnarray}
\Phi_2=
\left(
\begin{array}{cc}
\frac{1}{\sqrt{2}}\Phi_{-\frac{1}{3}}&\Phi_{\frac{2}{3}}\\
\Phi_{-\frac{4}{3}}&-\frac{1}{\sqrt{2}}\Phi_{-\frac{1}{3}}
\end{array}
\right).
\label{tr677}
\end{eqnarray}
with the covariant derivative:
\begin{eqnarray}
D_{\mu}\Phi_i=\partial_{\mu}\Phi_i+ig[W_{\mu},\Phi_i]+ig'YB_{\mu}\Phi_i.
\label{cobf6657}
\end{eqnarray}
The $Y=2$ Higgs triplet has the form \cite{Gunion}:
\begin{eqnarray}
\Phi_1=
\left(
\begin{array}{cc}
\frac{1}{\sqrt{2}}\Phi_1^{+}&-\Phi_1^{++}\\
\Phi_1^{0*}&-\frac{1}{\sqrt{2}}\Phi_1^+
\end{array}
\right).
\label{tr67744}
\end{eqnarray}
and a covariant derivative similar to that in Eq. (\ref{cobf6657}).
Finally the Higgs sextet has the structure: $\sigma=(\sigma_1,\sigma_2,\sigma_3,\sigma_4,\sigma_5,\sigma_6)^T$ where each $\sigma_j$ is an $SU(2)_Y$ Higgs triplet as shown above. Alternatively one can use the notation $\sigma_{ij}$ with two color symmetric indices. Of relevance is only the covariant derivative with respect to the $SU(3)_c$ group which has the form \cite{Brauner}:
\begin{eqnarray}
D_{\mu}\sigma_{ij}=\partial_{\mu}\sigma_{ij}-ig_c G^a_{\mu}(\frac{1}{2}\lambda^a_{ik}\sigma_{kj}+\sigma_{ik}\frac{1}{2}\lambda^2_{kj}).
\label{cob5664}
\end{eqnarray}
Yukawa interaction is given by ${\cal L}_4$;
\begin{eqnarray}
&&{\cal L}_4=f_i\bar{l}_{Li}H_i^{1}e_{Ri}+f'_i\bar{l}_{Li}\bar{H}_i^{1}\nu_{Ri}+y_i\bar{Q}_L\bar{H}_i^{2}u_{Ri}+y_i'\bar{Q}_L  H_i^{2}d_{Ri}+
\nonumber\\
&&+ih_{ij}^1l_{Li}^TC\sigma^2\Phi_1l_{Lj}+ih_{ij}^2Q_{Li}^TC\sigma^2(\Phi_2+\Sigma)q_{Lj}+h.c.
\label{res44355234}
\end{eqnarray}

Here $\bar{H}_i^j=i\sigma^2H_i^{*j}$ and the indices $i$, $j$ span the three generations.

The final term  in Eq. (\ref{res5534435}) is ${\cal L}_5$ which corresponds to the Higgses interactions.  There are many possible renormalizable terms especially in the sector of the Higgs doublets. Since the selection depends on many assumptions and choices we shall not discuss this term in detail here.  Of utmost importance is that the color Higgses have no spontaneous symmetry breaking so that the $SU(3)_c$ symmetry is preserved and that the multiplets interact in such a way  to permit the extraction of a single Higgs doublet with a  vev at the electroweak scale. Thus we expect that the masses of the additional Higgs particles are very large and the couplings with the standard model fermions are very small to be in accordance with the latest experimental data regarding the electroweak sector and the presence of additional Higgs particles.  The model might produce dark matter candidates among its multiple scalar states.

\section*{Acknowledgments} \vskip -.5cm

The work of R. J. was supported by a grant of the Ministry of National Education, CNCS-UEFISCDI, project number PN-II-ID-PCE-2012-4-0078.


\begin{thebibliography}{15}
\bibitem{Glashow} S. L. Glashow, Nuxl. Phys. {\bf 22} (4): 579-588 (1961).
\bibitem{Weinberg} S. Weinberg, Phys. Rev. Lett.  {\bf 19} (21):1264-1266 (1967).
\bibitem{Brout} F. Englert and R. Brout, Phys. Rev. Lett. {\bf 13} (16):321-323 (1964).
\bibitem{Higgs} P. W. Higgs, Phys. Rev. Lett.  {\bf 13}  (16): 508-509 (1964).
\bibitem{Salam} A. Salam, Elementary Particle Theory, Nobel Symposium No. 8, N. Svartholm (eds) (Almqvist and Wiksells, Stockholm 1968), p. 137.
\bibitem{Guralnik} G. S. Guralnik, C. R. Hagen and T. W. B. Kibble, Phys. Rev. Lett. {\bf 13}, 585 (1964).
\bibitem{Hooft} G. 't Hooft, Nucl. Phys. B {\bf 3}, 167 (1971).
\bibitem{Veltman} G. 't Hooft and M. J. G. Veltman, Nucl. Phys. B {\bf 44}, 189 (1972).
\bibitem{Jora} A. H. Fariborz and  R. Jora, arXiv:1412.7658 (2014).
\bibitem{Jora1} A. H. Fariborz and R. Jora, to appear.
\bibitem{Hagiwara} K. Hagiwara and J. Nakamura, JHEP 1302, 100 (2013).
\bibitem{Vecchi} L. Vecchi, JHEP 1110, 003 (2011), arXiv:1107.2933.
\bibitem{Dobrescu1} Y. Bai and B. A. Dobrescu, JHEP 1107, 100 (2011), arXiv:1012.5814.
\bibitem{Dobrescu2} B. A. Dobrescu and G. Z. Krnjaic,  Phys. Rev. D {\bf 85}, 075020 (2012), arXiv:1104.2893.
\bibitem{Georgi} H. Georgi and M. Machacek, Nucl. Phys. B {\bf 262}, 463-477 (1985).
\bibitem{Gunion} J. F. Gunion, R. Vega and J. Wudka, Phys. Rev. D {\bf 42}, 1673 (1990).
\bibitem{Brauner} T. Brauner, J. Hosek and R. Sykora, Phys. Rev. D {\bf 68}, 094004 (2003), arXiv:hep-ph/0303230.
\end{thebibliography}
\end{document}